\documentclass{article} 
\usepackage{iclr2021_conference,times}
\usepackage{caption}
\usepackage{graphicx}
\usepackage{booktabs}
\usepackage{bbm}
\usepackage{flushend}
\usepackage{subcaption}
\usepackage{amsmath}
\usepackage{amsthm}
\usepackage{mathrsfs}
\usepackage{mathtools}
\usepackage{soul}
\usepackage{tabularx}
\usepackage{paralist}

\usepackage{amsmath,amsfonts,bm}

\def\eqref#1{equation~\ref{#1}}

\def\1{\bm{1}}

\DeclareMathAlphabet{\mathsfit}{\encodingdefault}{\sfdefault}{m}{sl}
\SetMathAlphabet{\mathsfit}{bold}{\encodingdefault}{\sfdefault}{bx}{n}

\newcommand{\R}{\mathbb{R}}

\DeclareMathOperator*{\argmin}{arg\,min}

\usepackage{hyperref}
\usepackage{url}
\usepackage{float}
\usepackage{lineno}
\usepackage{tikz}
\usepackage{cleveref}
\usepackage{array}

\def\checkmark{\tikz\fill[scale=0.4](0,.35) -- (.25,0) -- (1,.7) -- (.25,.15) -- cycle;}
\title{Interpretability of Epidemiological Models: The Curse of Non-Identifiability}

\author{Ayush Deva, Siddhant Shingi, Avtansh Tiwari, Nayana Bannur, Sansiddh Jain, \\ \textbf{Jerome White, Alpan Raval, Srujana Merugu} \\
Wadhwani Institute for Artificial Intelligence\\
}

\newcommand{\ar}[1]{{\ifdraft\color{red}[Alpan: {#1}]\fi}}
\newcommand{\sj}[1]{{\ifdraft\color{blue}[Sansiddh: {#1}]\fi}}
\newcommand{\sm}[1]{{\ifdraft\color{blue}[Srujana: {#1}]\fi}}

\newcommand{\ad}[1]{{\ifdraft\color{violet}[Ayush: {#1}]\fi}}
\newcommand{\sid}[1]{{\ifdraft\color{orange}[Siddhant: {#1}]\fi}}
\newcommand {\commentout}[1] {}

\newcommand{\btheta}[0]{\boldsymbol{\theta}}
\newcommand{\bTheta}[0]{\boldsymbol{\Theta}}
\newcommand{\bx}[0]{\boldsymbol{x}}
\newcommand{\by}[0]{\boldsymbol{y}}

\newcommand{\bxdot}[0]{\boldsymbol{\dot{x}}}
\newcommand\Sdot{\dot{S}}

\newcommand{\minisection}[1]{#1.}

\iclrfinalcopy 
\begin{document}

\maketitle

\begin{abstract}
Interpretability of epidemiological models is a key consideration, especially when these models are used in a public health setting. Interpretability is strongly linked to the \emph{identifiability} of the underlying model parameters, i.e., the ability to estimate parameter values with high confidence given observations. In this paper, we define three separate notions of identifiability that
explore the different roles played by the model definition, the loss function, the fitting methodology, and the quality and quantity of data. We define an epidemiological compartmental model framework in which we highlight these non-identifiability issues and their mitigation.





\end{abstract}


\section{Introduction} 
The global COVID-19 pandemic has spurred intense interest in epidemiological compartmental models~\citep{thompson2020epidemiological, brauer2008compartmental}. The use of epidemiological models is driven by four main criteria:  \emph{expressivity} to faithfully capture the disease dynamics; \emph{learnability} of parameters conditioned on the available data; 
\emph{interpretability} to understand the evolution of the pandemic; and \emph{generalizability} to future scenarios by incorporating additional information.

Compartmental models are popular because they are relatively simple and known to be highly expressive and generalizable. However, their interpretability and learnability 
depend strongly on the alignment between  data observations and model complexity.  Different choices of model parameters can often lead to (approximately) the same forecast case counts, leading to what is commonly referred to as non-identifiability \citep{raue2009structural,jacquez1990parameter}. The problem of non-identifiability is only exacerbated with increased model complexity. 

This lack of identifiability is detrimental because
\begin{inparaenum}[(a)]
\item parameter distributions estimated from the observed data tend to be biased with large variances, thus precluding easy interpretation, and  
\item non-identifiable models typically have reduced accuracy on long-time forecasts due to the high parameter variance. 
\end{inparaenum}
This phenomenon is illustrated in \autoref{fig:test_period}, where the forecasting errors between a non-identifiable model and its reparametrized version (later shown to be identifiable) are compared.

Non-identifiability in epidemiological models is rooted in the model dynamics, in the fitting loss function and methodology, and in the quality and  quantity of data available. Identifiability is typically broadly classified into structural (i.e., purely model-dependent)~\citep{reid1977structural,massonis2020structural} and practical  (i.e., data-, loss- and fitting methodology-dependent), with the latter 
often defined vaguely, and in the context of specific loss functions \citep{raue2009structural,wieland2021structural}.

\minisection{\textbf{Contributions}} 
This paper delineates various general notions of model identifiability that are contextualized in compartmental epidemiological models,
including a novel notion of \emph{statistical identifiability} that depends on the loss function optimized in estimation, and an empirical framework to assess  
practical identifiability in terms of the highest posterior density intervals. 
We study these ideas in the specific context of a SEIR-like compartmental model (SEIARD) that was  deployed in a major densely populated city in India for case forecasting to inform capacity and policy decisions.

\begin{minipage}[t]{\linewidth}
  \begin{minipage}{0.38\linewidth}
    \includegraphics[width=\linewidth]{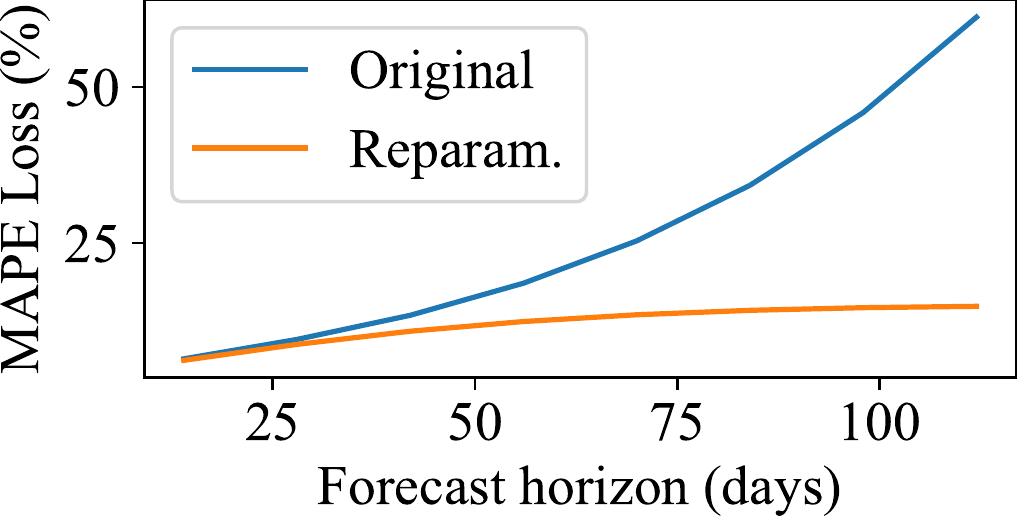}
    \captionof{figure}{Forecast error. \autoref{sec:nonid} defines ``Original'' and ''Reparam.''}
    \label{fig:test_period}
  \end{minipage}%
  \hfill%
  \begin{minipage}{0.59\linewidth}
    \small
    \begin{tabularx}{\linewidth}{Xccc}
      \toprule
      & \textbf{Structural}
      & \textbf{Statistical}
      & \textbf{Practical} \\
      \midrule
      Model form & \checkmark & \checkmark & \checkmark \\
      Loss function & & \checkmark & \checkmark \\
      Observation interval & & \checkmark & \checkmark \\
      Noisy data & & & \checkmark \\
      Fitting method & & & \checkmark \\
      \bottomrule
    \end{tabularx}
    \captionof{table}{Notions of identifiability.}
    \label{fig:table}
  \end{minipage}
\end{minipage}


\section{Notions of Identifiability} \label{sec:notions}
Consider a dynamical system $M$ characterized by  $\bxdot = f(\bx, \btheta)$,  where $\bx$ is the system state,
 $\bxdot$ the time derivative,  and $\btheta$  the 
 parameters. Let $\by(t)=g(\bx(t),\btheta)$
 be the observation function that maps   state $\bx$ to observations at time $t$. Combining these, 
 we may  express $\by(t) = h(\bx_0, \btheta,t)$, where  $\bx_0$ is the initial state. 
Epidemiological compartmental models are 
    dynamical systems in which states correspond to  compartmental population counts, with a subset or aggregates of these counts being observed. We consider three notions of identifiability for such a system~(\autoref{fig:table}).

\noindent \textbf{Structural Identifiability.} 
We say that a parameter $\theta_i$ in a model M 
is structurally identifiable if for any $\btheta$ in the domain,
$h(\hat{x}_0, \hat{\btheta},t)  = h(x_0, \btheta,t)\,\, \forall t\implies \hat{\theta}_i = \theta_i$, i.e., distinct parameter choices result in distinct observation series.
Note that in partially observable systems, this notion of identifiability also applies to components of the initial state $x_0$ that are not observed.
In linear time-invariant systems (LTI), where  $\bxdot = B(\btheta)\bx$ and  $\by = C(\btheta)\bx$,  the resulting solution takes the form  $\bx = e^{B(\theta)t} \bx_0$. Structural identifiability for this case is characterized in terms of the properties of $B(\btheta)$ and $C(\btheta)$~\citep{kalman1959general, sontag2013mathematical}.  
Recent works~\citep{martinelli2020rank, villaverde2019observability, massonis2020structural}  extend these  results  to nonlinear systems and SEIR model variants.

\noindent\textbf{Statistical Identifiability.} We propose a new notion of identifiability that depends both on  the parametric form of the model and the statistical estimation process.  
Let $y_M(t) = h(\bx_0, \btheta,t)$ denote the model output and  $y_D(t)$  the  observations over a finite time horizon $t \in [t_b,t_e]$. Parameter estimation given the observations is carried out by optimizing a loss function $\mathscr{L}(\cdot)$, i.e., 
   $ \btheta^* =  \argmin_{\btheta \in \bTheta} \mathscr{L}_D(\btheta) = \argmin_{\btheta \in \bTheta} \sum_{t=t_b}^{t_e} \mathscr{L}\left(y_D(t), y_M(t)\right).$
We consider a parameter-wise loss function, called the \emph{profile-likelihood} (PL) of $\theta_i$ \citep{raue2009structural}, defined as  
where $\btheta_{-i}= \btheta\setminus\theta_{i}$ denotes the complementary set of parameters.
We say a parameter $\theta_i$ is \textit{statistically identifiable} within a domain $\Theta_i$ if $\mathscr{L}_D^{i}(\theta_i)$ is strictly convex with respect to $\theta_i \in \Theta_i$. Though this ensures a unique global optimum, it does not necessarily imply joint convexity of $\mathscr{L}_D$ with respect to $\btheta$. 

Clearly,  statistical identifiability implies  structural identifiability, but the converse is not always true. For example, a loss function with a flat extended minimum will lead to lack of statistical identifiability even though the model itself is structurally identifiable.
However, structural identifiability does imply statistical identifiability for a LTI system with a strictly convex loss function, 
when the number of  observations at distinct times in $D$ is at least equal to the 
rank of the observability  matrix~\citep{villaverde2019observability,kalman1959general}. 
The proof follows from the strict convexity of the loss function, the convexity of the solution $e^{B(\theta)t}\bx_0$ itself, and the fact that  the composition of a convex function with a convex non-decreasing function remains convex~\citep{boyd2004convex}.


\textbf{Practical Identifiability Intervals.} Models could be both structurally and statistically identifiable, but parameter fitting could nevertheless present practical challenges, related to application-specific tolerance to error on the fitted parameters~\citep{roosa2019assessing}. This situation is captured through the notion of \emph{practical identifiability}. 
Given a model $M$, data $D$, loss function $\mathscr{L}_D$, and fitting method $H$, let $p_D(\theta_i)$ be the resulting posterior distribution of $\theta_i$.  
Following ~\cite{raue2009structural}, we 
characterize practical identifiability in terms of the $\alpha$-identifiability intervals of the parameters, in two different ways.

The first approach defines this interval  in terms of level sets of the loss function, i.e., 
$J_{PL}^i(\alpha,\mathscr{L}_D, H) = \{\theta_i  | \mathscr{L}_D^i(\theta_i)  \leq  \mathscr{L}_D^{\alpha}\},$ where $\mathscr{L}_D^{\alpha}$ is the $\alpha$-quantile of $\mathscr{L}_D(\cdot)$, defined as the level set with probability mass $\alpha$ as given by the posterior distribution. The special case of the squared loss where $\mathscr{L}_D^{\alpha}$ is the sum of $\mathscr{L}_D(\theta^*)$ and the $\alpha$-quantile of  $\chi^2$ distribution with a single degree of freedom is discussed in ~\citep{wieland2021structural,raue2013joining}. 

The second, Bayesian approach defines the $\alpha$-identifiability interval  $J_{PP}^i(\alpha, \mathscr{L}_D, H)$ as a \textit{highest posterior density interval} (HPDI)~\citep{wright1986note} with probability mass  $\alpha$. When $\mathscr{L}_D^i(\cdot)$ is strictly convex in $\theta_i$ and the marginal posterior distribution $p_D^i(\theta_i)$ is monotonically decreasing with respect to  $\mathscr{L}_D^i(\cdot)$, the two definitions can be shown
to be equivalent based on  the convexity properties and one-one association of the corresponding level sets~\citep{boyd2004convex}. Empirical analysis in~\cite{raue2013joining} suggests a similar relationship. See~\autoref{sec:A4} for a detailed analysis.

\section{Non-identifiability in Epidemiological Models} 
\label{sec:nonid}
\begin{figure}[t]
\begin{minipage}[t]{0.50\linewidth}
    \includegraphics[width=\linewidth]{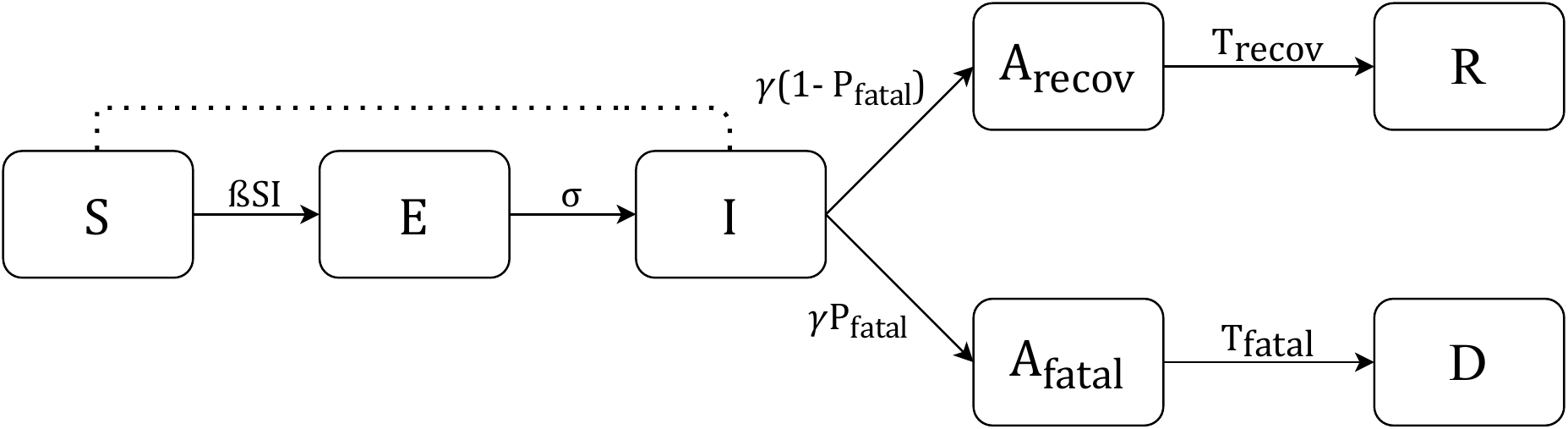}
    \caption{SEIARD Model.}
    \label{fig:model_v1}
\end{minipage}%
    \hfill%
\begin{minipage}[t]{0.45\linewidth}
    \includegraphics[width=\linewidth]{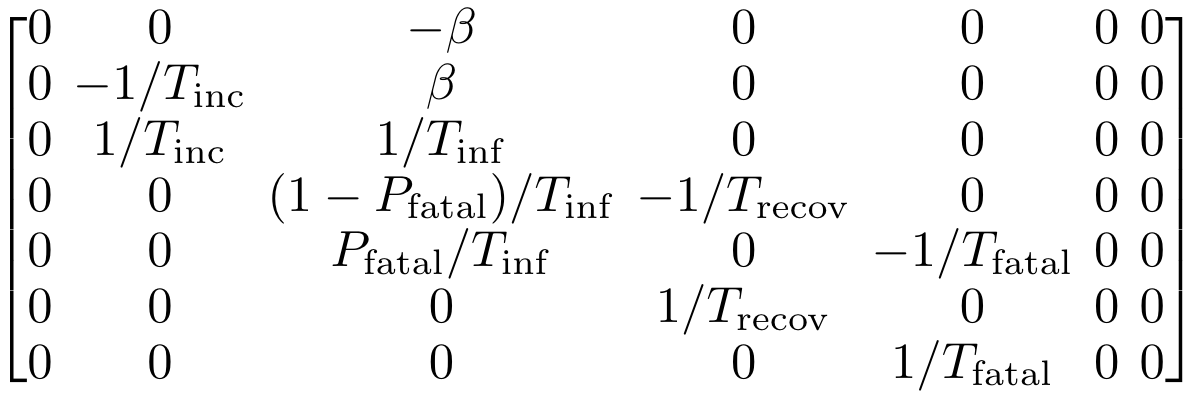}
    \caption{SEIARD Dynamics Matrix (LTI when $S\simeq N$) }
    \label{fig:B_theta}
\end{minipage}
\end{figure}

\textbf{SEIARD Model.}  \autoref{fig:model_v1} shows the SEIARD model, which is an extension of the well-known SEIR model~\citep{Hethcote}.
The $\mathbf{S}$ (Susceptible), $\mathbf{E}$ (Exposed), and $\mathbf{I}$ (Infectious) compartments and their associated parameters (transmission rate $\beta$, incubation period $T_{\rm inc} = \sigma^{-1}$, and  infectious period $T_{\rm inf} = \gamma^{-1}$) are defined as in the  SEIR model.  The post-infectious stage consists of parallel paths through $\mathbf{A}_{\rm fatal}$  or $\mathbf{A}_{\rm recov}$, depending on the  eventual outcome: fatality ($\mathbf{D}$) with probability $P_{\rm fatal}$ or recovery ($\mathbf{R}$), with $T_{\rm fatal}$ and $T_{\rm recov}$ being the respective time durations. Observed case counts are  mapped to compartment populations as follows: recovered cases ($R$)$ = |\mathbf{R}| $, deceased cases ($D$)$ =|\mathbf{D}| $, active cases ($A$)$ = |\mathbf{A}_{\rm recov}| + |\mathbf{A}_{\rm fatal}|$.
Defining state $\bx = (S,E,I, A_{\rm recov}, A_{\rm fatal}, R,D)$  and observations $\by = (A_{\rm recov}+A_{\rm fatal}, R,D)$, the model
dynamics~(\autoref{fig:B_theta}) reduces to that of an LTI~\citep{kalman1959general} in the early stages of the epidemic, because one can then set
$S \simeq N$ to first approximation, thus yielding $\Sdot = - \beta SI/N \simeq - \beta I$.


\noindent \textbf{Parameter Estimation.} In our experiments, synthetic data was simulated from the SEIARD model using realistic parameters (\autoref{sec:A3}). 
We treat the initial state values of \textbf{E} ($E_0$) and \textbf{I} ($I_0$) compartments as unobserved parameters.
We estimate model parameters (including $E_0$ and $I_0$) by optimizing the mean absolute percentage error (MAPE) between the true and predicted values for each of the three observed time series in $\boldsymbol{y}$ as well as that of the total count (sum of these three).
Fitting is performed using two different approaches: Sequential Model Based Optimization (SMBO) based on Tree-structured Parzen Estimators as implemented in the HyperOpt library~\citep{hyperopt}, and an  MCMC-based sampling approach (see \autoref{mcmc_details}). 

\begin{figure}
    \centering
    
    \begin{subfigure}[b]{.31\linewidth}
        \includegraphics[width=\linewidth]{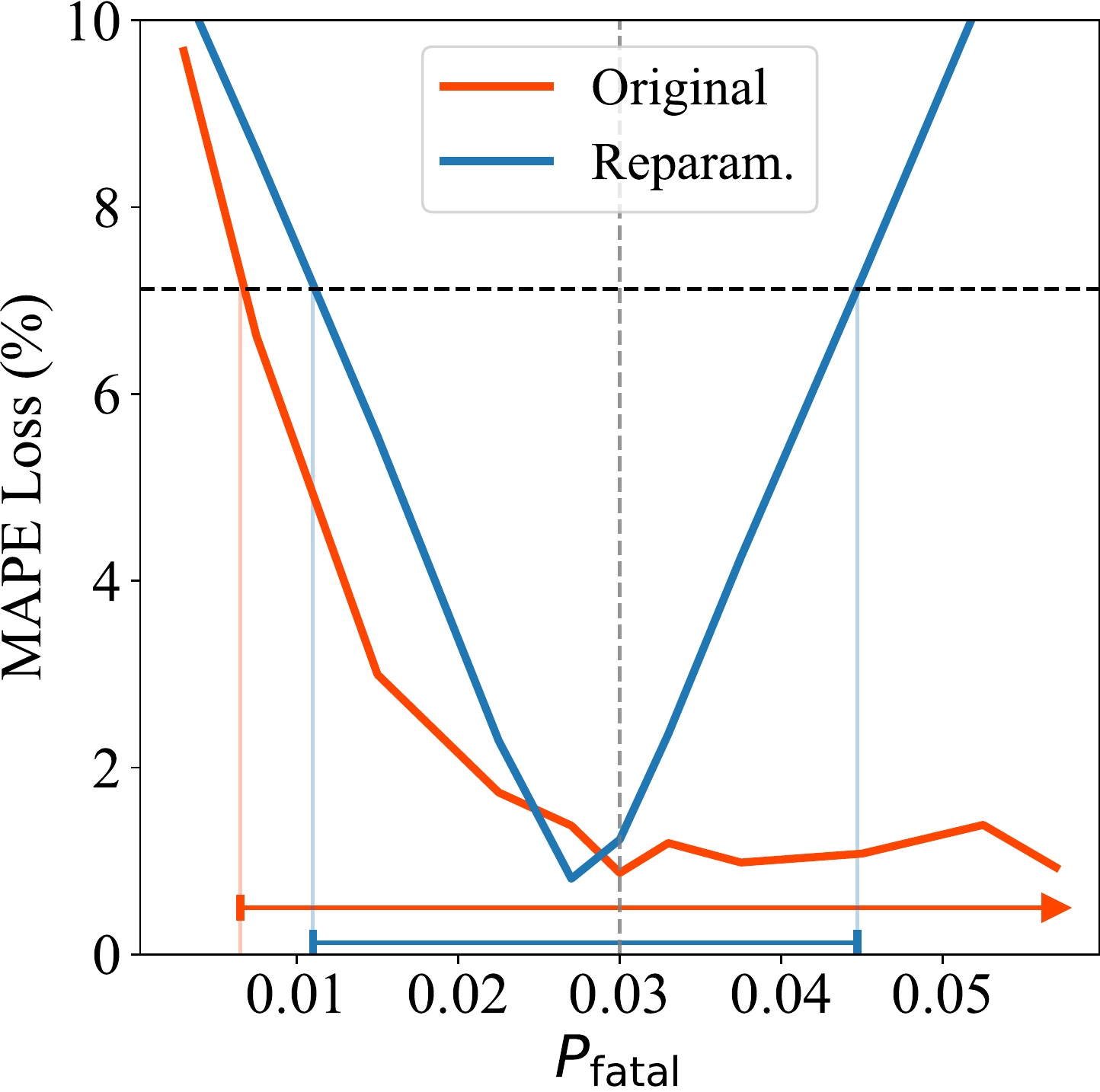}
        \caption{}\label{fig:pl_p_fatal}
    \end{subfigure}
    \hspace{2mm}
    \begin{subfigure}[b]{.31\linewidth}
        \includegraphics[width=\linewidth]{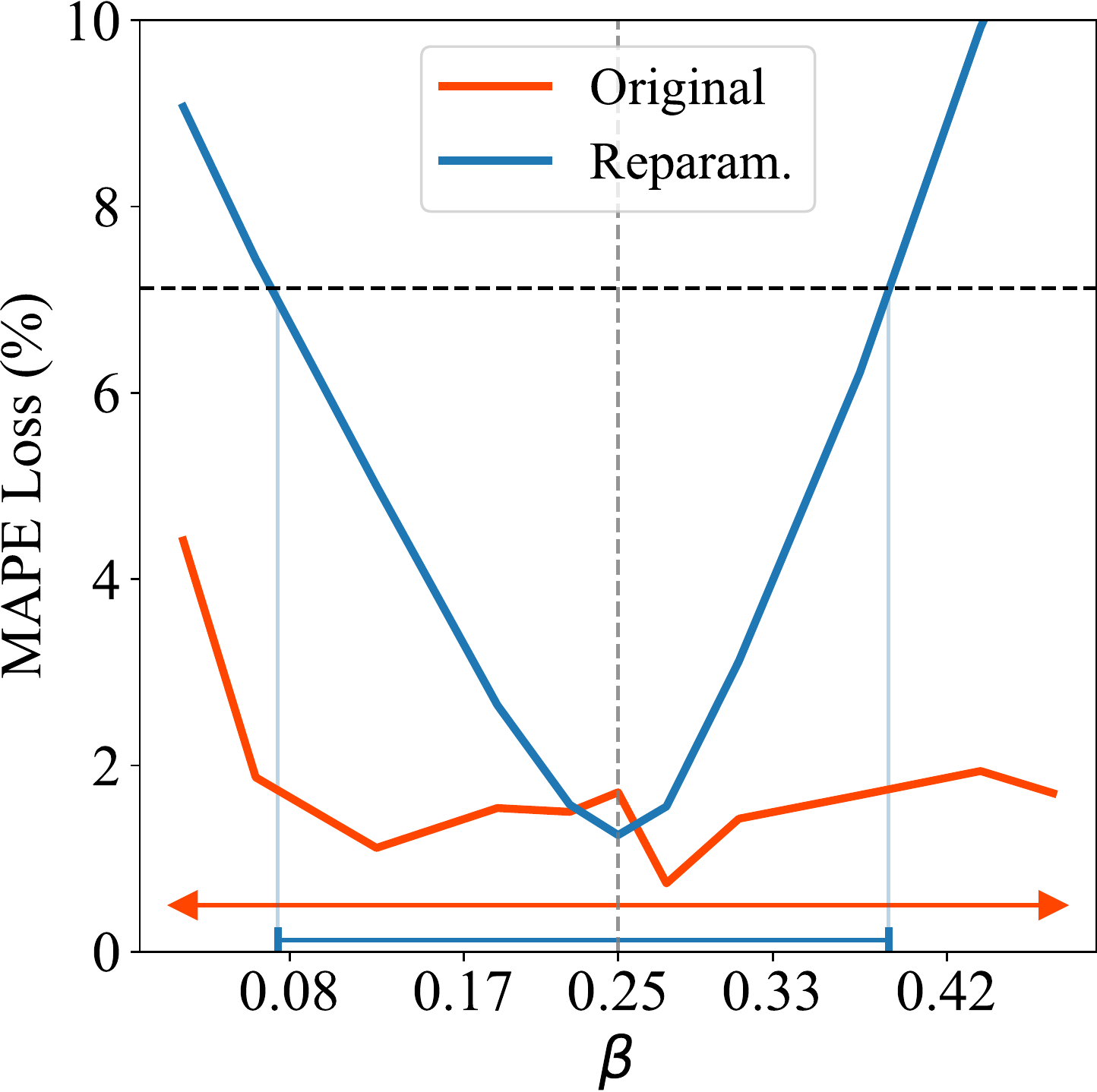}
        \caption{}\label{fig:pl_beta}
    \end{subfigure}
    \hspace{2mm}
    \begin{subfigure}[b]{.31\linewidth}
        \includegraphics[width=\linewidth]{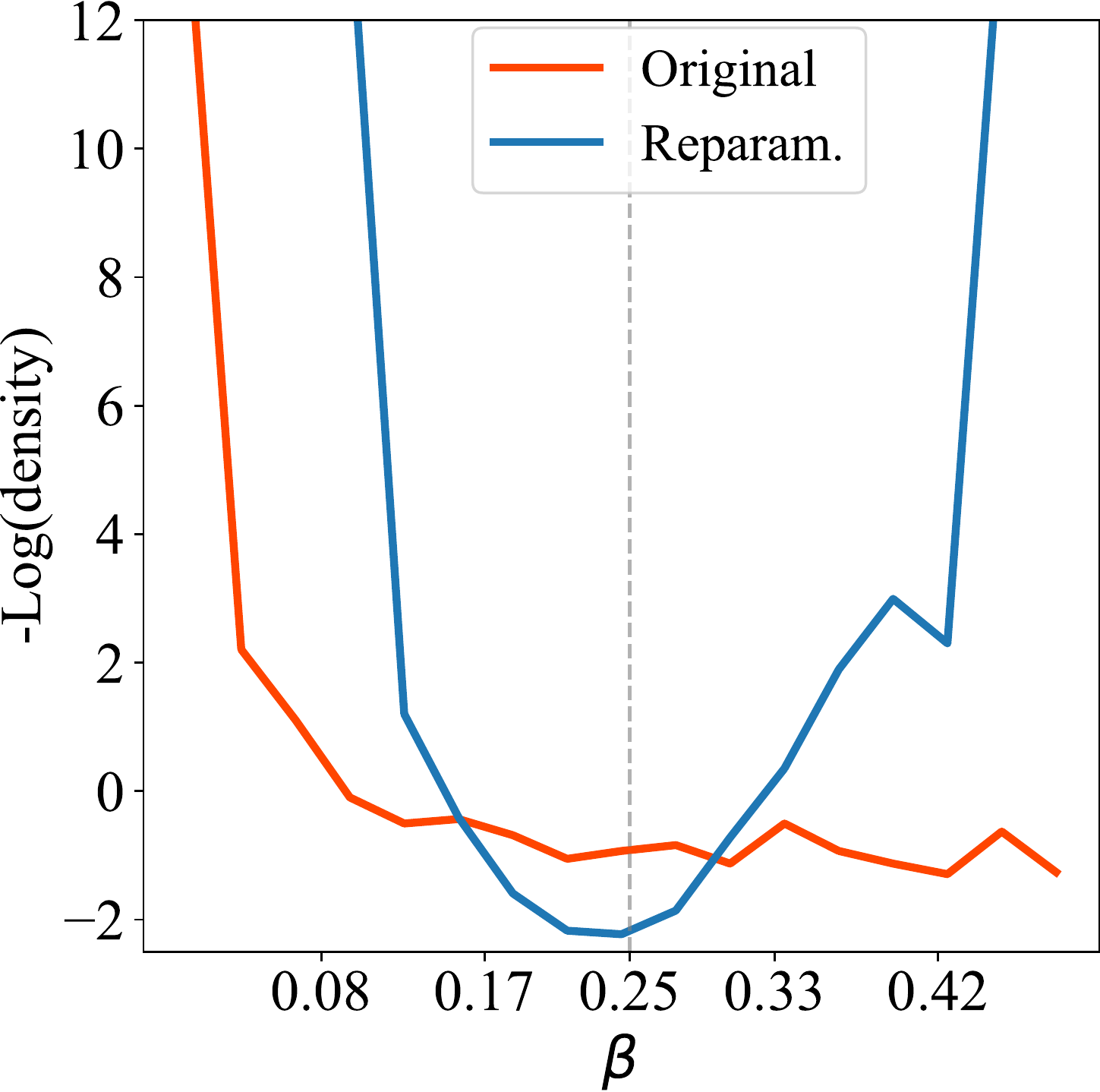}
        \caption{}\label{fig:mcmc_combine}
    \end{subfigure}
    
    \begin{subfigure}[b]{.31\linewidth}
        \includegraphics[width=\linewidth]{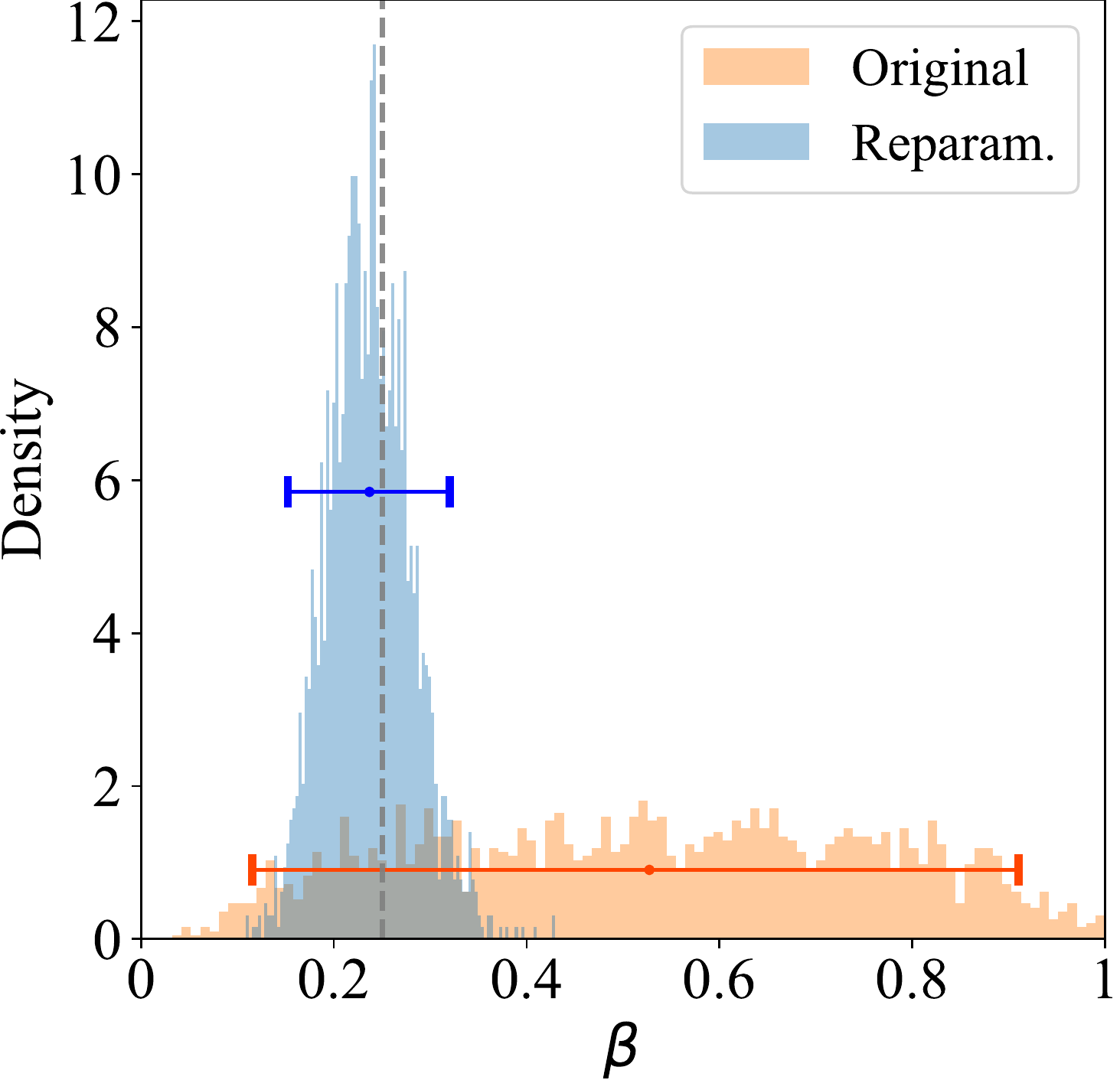}
        \caption{}\label{fig:mcmc_dist}
    \end{subfigure}
    \hspace{2mm}
    \begin{subfigure}[b]{.31\linewidth}
        \includegraphics[width=\linewidth]{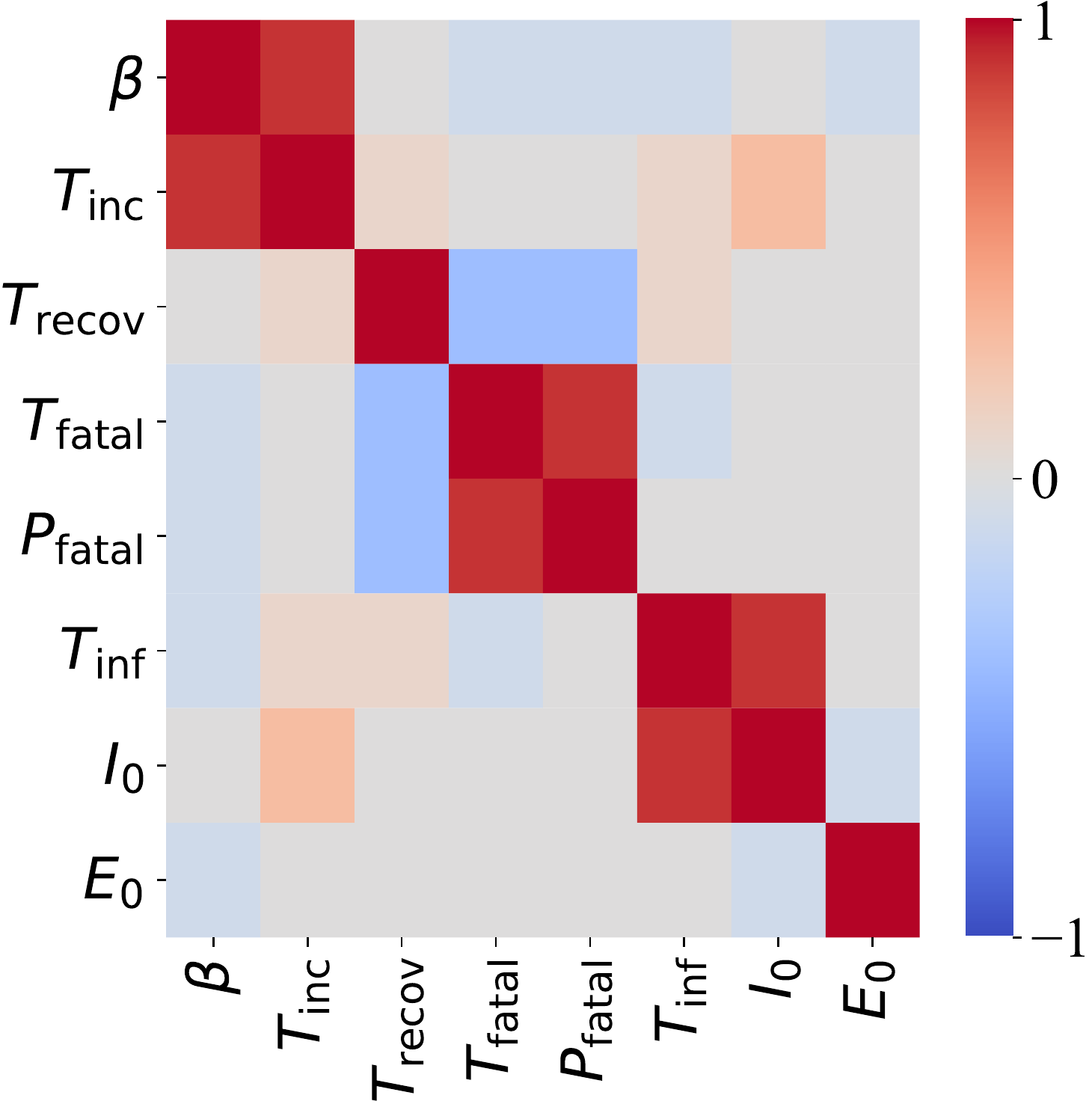}
        \caption{}\label{fig:mcmc_corr}
    \end{subfigure}
    \hspace{2mm}
    \begin{subfigure}[b]{.31\linewidth}
        \includegraphics[width=\linewidth]{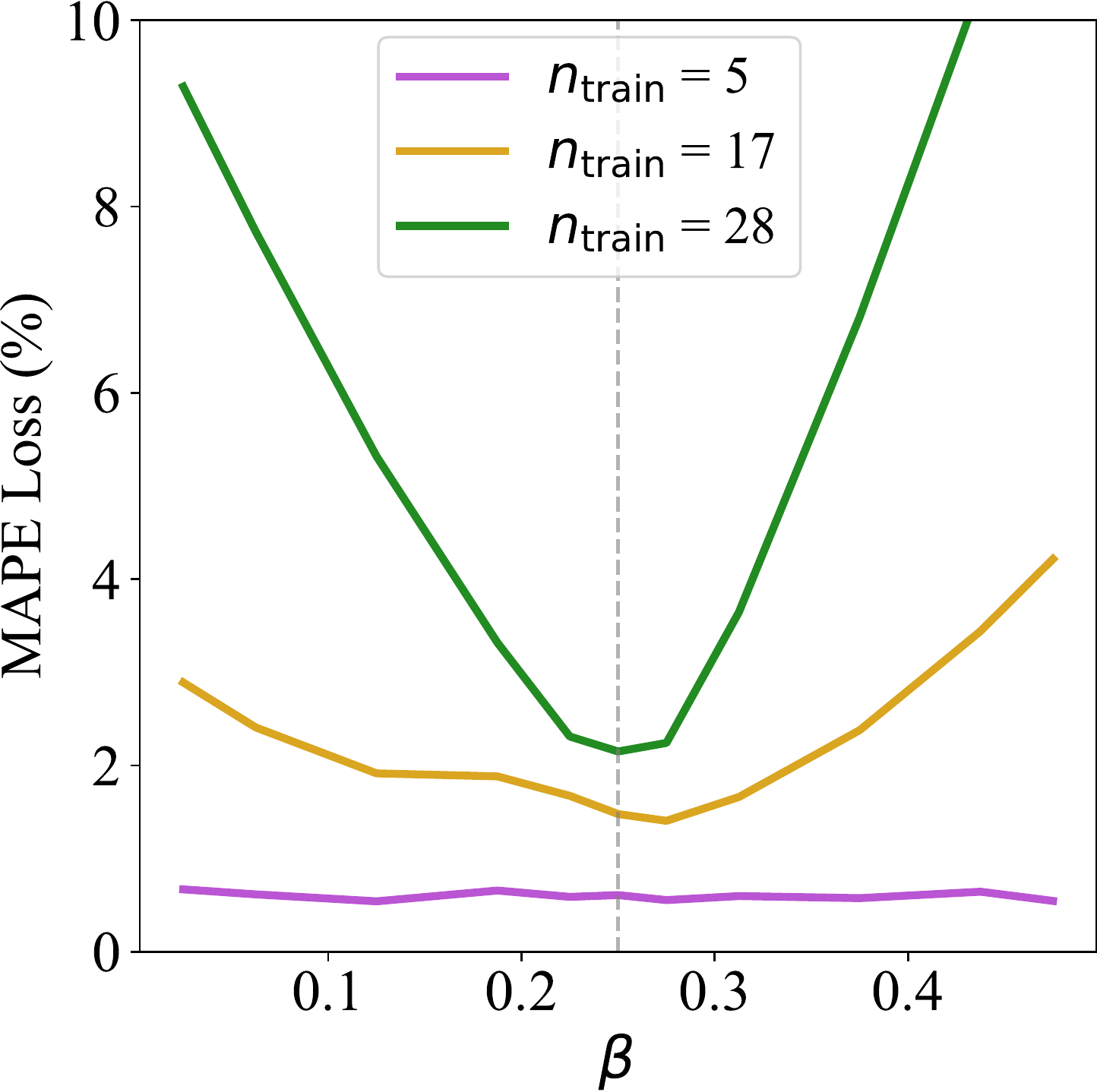}
        \caption{}\label{fig:pl_training}
    \end{subfigure}
    
    \caption{Panels~\subref{fig:pl_p_fatal} and~\subref{fig:pl_beta} show  PL curves for $P_{\rm fatal}$ and $\beta$ respectively. The horizontal black line depicts the $95\%$ quantile loss as defined in \autoref{sec:notions}, while the vertical gray lines are the true values given in \autoref{sec:A3}. Error bars indicate  confidence intervals based on where the curves intersect the black line; an arrow head indicates that the interval extends beyond the plotted range. Panel~\subref{fig:mcmc_combine} shows the negative logarithm of the posterior density of $\beta$; the corresponding posterior density is shown in panel~\subref{fig:mcmc_dist}. Error bars in~\subref{fig:mcmc_dist} refer to the $95\%$-HPDI (\autoref{sec:notions}) of $\beta$. Panel~\subref{fig:mcmc_corr} is the correlation matrix of parameter samples obtained via MCMC sampling for the Original model. Panel~\subref{fig:pl_training} shows PL curves of $\beta$ for the Reparam. model for different training durations.}
    \label{fig:figure_2}
\end{figure}

\noindent \textbf{Identifiability in SEIARD Model.} Structural identifiability analysis similar to that in \cite{massonis2020structural} indicates that the SEIARD model is structurally non-identifiable. 
This may be attributed to  the unobserved compartments ($E$, $I$) as well as the split between $A_{\rm recov}$ and $A_{\rm fatal}$. A larger fatality rate  with a larger delay $T_{\rm fatal}$ would be indistinguishable from a smaller fatality with a smaller delay just by observing the active, deceased and recovery counts. This structural non-identifiability also manifests as \emph{statistical non-identifiability}, as can be seen in the shape of the profile likelihood curves corresponding to the non-reparametrized   (Original) SEIARD model in \autoref{fig:pl_p_fatal} and~\ref{fig:pl_beta}.
Similarly, the confidence intervals of the parameters of this model, as shown in \autoref{fig:pl_p_fatal},~\ref{fig:pl_beta}, and~\ref{fig:mcmc_dist}, are broad, indicating considerable \emph{practical non-identifiability}.

\noindent \textbf{Improving Identifiability in SEIARD Model.} 
Non-identifiability in dynamical systems can often be attributed to correlated parameters. 
\autoref{fig:mcmc_corr}  shows the correlation between parameters constructed using  samples from the joint posterior distribution. A common way ~\citep{joubert2020efficient} to reduce non-identifiability is thus through reparameterization of models by constraining a parameter subset. 
The choice of parameters to constrain may be informed by secondary information sources. 
For epidemiological models, parameters such as the incubation period $T_{\rm inc}$, infectious period $T_{\rm inf}$, and the time to death $T_{\rm fatal}$ can be estimated from medical literature or line lists of representative case data. A model where these parameters
are fixed to their true values (see \autoref{sec:A3}) is referred to as the reparameterized (Reparam.) SEIARD model, which can be shown to be structurally identifiable~\citep{massonis2020structural}. 
The PL curves for the reparameterized model in \autoref{fig:pl_p_fatal} and~\ref{fig:pl_beta}, and the posterior density plot in \autoref{fig:mcmc_dist} 
 have relatively tighter confidence intervals, indicating an improvement in practical identifiability. Additionally, 
 \autoref{fig:mcmc_combine} empirically indicates that the PL curves and the negative logarithm of the posterior density give similar information.


\noindent \textbf{Data Dependence of Identifiability}
\autoref{fig:pl_training} shows the PL curves for the reparameterized model with various training durations, 
indicating increased practical identifiability with more observations. 
A larger set of observations may therefore serve to further fine-tune parameters in such a setting.

\section{Conclusions and Future Work}
We presented a novel framework to analyze identifiability of model parameters in epidemiological models, using the various lenses of model structure, loss functions and fitting methodology, and data. While the notions of structural and statistical identifiability are useful for detecting and resolving non-identifiability in an analytic fashion via reparameterization, practical identifiability intervals  function as an effective tool for fine-tuning the parameter estimation process. We present these ideas and empirical results in the specific context of the SEIARD compartmental model. In the future, we plan to
explore connections between different types of identifiability and empirically analyze the identifiability of multiple SEIR variants on real and synthetic data.

\subsubsection*{Acknowledgments}
\small{
This study is made possible by the generous support of the American People through the United States Agency for International Development (USAID). The work described in this article was implemented under the TRACETB Project, managed by WIAI under the terms of Cooperative Agreement Number 72038620CA00006. The contents of this manuscript are the sole responsibility of the authors and do not necessarily reflect the views of USAID or the United States Government. This work is co-funded by the Bill and Melinda Gates Foundation, Fondation Botnar and CSIR - Institute of Genomics and Integrative Biology.

This work was enabled by several partners through the informal COVID19 data science consortium, most notably a group of volunteer data scientists under the umbrella group ‘Data Science India vs. COVID’.
We would like to thank the Smart Cities Mission for providing early impetus to this work, the Brihanmumbai Municipal Corporation and the Integrated Disease Surveillance Program (Jharkhand) for their support.
}




\bibliography{iclr2021_conference}
\bibliographystyle{iclr2021_conference}

\newpage
\appendix
\section{Appendix}

\subsection{SEIARD Model Dynamics}
\label{sec:A1}
The dynamical equations governing the transitions in SEIARD model (\autoref{fig:model_v1}) are
\begin{align}
    \frac{dS}{dt} &=-\beta\frac{IS}{N}; \;
    \frac{dE}{dt} =\beta\frac{IS}{N} - \sigma E; \;
    \frac{dI}{dt} =\sigma E - \gamma I,\\
    \frac{dA_{\rm recov}}{dt} &= \left(1 - P_{\rm fatal}\right) \cdot \gamma I - \frac{A_{\rm recov}}{T_{\rm recov}}, \\
    \frac{dA_{\rm fatal}}{dt} &= P_{\rm fatal} \cdot \gamma I - \frac{A_{\rm fatal}}{T_{\rm fatal}}, \\
    \frac{dR}{dt} &= \frac{A_{\rm recov}}{T_{\rm recov}}; \;
    \frac{dD}{dt} = \frac{A_{\rm fatal}}{T_{\rm fatal}},
\end{align}
where $N$ is the city population, $\beta$, $\sigma$, and $\gamma$ are the standard epidemiological parameters for a SEIR model, $P_{\rm fatal}$ is the transition probability to the mortality branch, and $T_{\rm recov}$ and $T_{\rm fatal}$ are timescales that govern transitions out of the $\mathbf{A}_{\rm recov}$ and $\mathbf{A}_{\rm fatal}$ compartments. Variables $S$, $E$, $I$, $A_{\rm recov}$, $A_{\rm fatal}$, $R$, $D$ denote the populations of the similarly named compartments. 

In the early stages of the epidemic, when $S \simeq N$, the model dynamics reduces to that of an LTI, with $\bxdot = B(\btheta)\bx$ and  $\by = C(\btheta)\bx$. For our model, we have $B(\btheta)$ and $C(\btheta)$ as shown below:

\[
\setlength\arraycolsep{2pt}
B(\btheta) = 
\begin{bmatrix}
    0 & 0 & -\beta & 0 & 0 & 0 & 0\\
    0 & -1/T_{\rm inc} & \beta & 0 & 0 & 0 & 0\\
    0 & 1/T_{\rm inc} & 1/T_{\rm inf} & 0 & 0 & 0 & 0\\
    0 & 0 & (1-P_{\rm fatal})/T_{\rm inf} & -1/T_{\rm recov} & 0 & 0 & 0\\
    0 & 0 & P_{\rm fatal}/T_{\rm inf} & 0 & -1/T_{\rm fatal} & 0 & 0\\
    0 & 0 & 0 & 1/T_{\rm recov} & 0 & 0 & 0\\
    0 & 0 & 0 & 0 & 1/T_{\rm fatal} & 0 & 0
\end{bmatrix}
,C(\btheta) = 
\begin{bmatrix}
    0 & 0 & 0 & 1 & 1 & 0 & 0\\
    0 & 0 & 0 & 0 & 0 & 1 & 0\\
    0 & 0 & 0 & 0 & 0 & 0 & 1
\end{bmatrix}
\]

\subsection{MCMC Implementation Details} \label{mcmc_details}
 Let  $X[t] = [X_h[t]]_{h \in \mathcal{H}}$ be a multivariate time series with $X_h[t]$ denoting the $h^{\text{th}}$ compartment time-series, and  $\mathcal{H}$ be the set of indices of components. Let the fitting period be given by $[t_i,t_j]$. The key components of our MCMC-within-Gibbs sampling are as follows.

\minisection{\textbf{Likelihood function}} 
We assume  
a likelihood function of the form 
\begin{equation}
\label{eq:likelihood}
 P(X \mid \btheta, s) = \prod_{h \in  \mathcal{H}} \prod_{t=t_i +1}^{t_j} \mathcal{N}(\hat{z}_{h,\btheta}[t] \mid z_h[t],s),  \end{equation}
where $\mathcal{N}(z \mid \mu, \sigma^2)$ denotes the Normal distribution pdf with mean $\mu$ and variance $\sigma^2$ following an appropriate conjugate prior. Further, 
    $z_h[t] = \log(X_h[t]) - \log(X_h[t - 1]),$
and $\hat{z}_{h,\btheta}[t]$ is the forecast equivalent of $z_h[t]$ and $s$ is the variance of the normal likelihood function, which is discussed later in this section. 

\minisection{\textbf{Proposal distribution}} At iteration $k$, we generate the samples from the proposal distribution for accept-reject step as
\begin{equation}
    \label{proposal-dist}
    \btheta \sim Q(\btheta_{k-1}, \Sigma_{\text{prop}}, \btheta_{\min}, \btheta_{\max})
\end{equation}
where $\btheta_{k-1}$ is the parameter vector chosen at $k-1$, and $Q(\cdot)$ is the pdf of a multivariate truncated Gaussian with the parameter range $[\btheta_{\min},\btheta_{\max}]$ and covariance matrix $\Sigma_{\text{prop}}$ as listed in \autoref{tab:param-ranges} (MCMC Proposal Variance).

We further assume that $s$ has the conjugate prior $s \sim {\rm InvGamma}(u,v)$ where $u = 40 $ and $v = 2/700 $ are hyperparameters.
Thus, it is straightforward to show, by multiplying the Normal likelihood with the prior, that if the MCMC chain has sampled parameters $\btheta_k$, the sample $s_k \sim P(s_k \mid \btheta_k, X)$ is also drawn from an Inverse Gamma distribution with parameters
        \begin{equation*}
        u_k = u + 2(t_j - t_i - 1), \;\;
        v_k = v + \sum_{h \in  \mathcal{H}}\sum_{t=t_i +1}^{t_j} \frac{(z_h[t] - \hat{z}_{h,\btheta_k}[t])^2}{2}.
       \end{equation*}
\newpage

\subsection{Hyperparameters} \label{sec:A3}

\autoref{tab:param-ranges} contains the parameter ranges and values used for all experiments in \autoref{sec:nonid}. The first row enumerates the  true value of parameters used to simulate the synthetic dataset while the second lists the search space for each parameter while fitting. The proposal variance used in \autoref{proposal-dist} is listed in the bottom row of \autoref{tab:param-ranges}. \autoref{tab:param-ranges-hyper} contains all the hyperparameters used to generate the synthetic dataset. ${\rm R}_{\rm 0}$, ${\rm A}_{\rm 0}$, ${\rm D}_{\rm 0}$ denote the initial values of recovered, active and deceased cases and N denotes total population.
\newcommand{\interval}[2]{[#1, #2]}

\newcommand{\newinterval}[2]{[#1, #2]}

\begin{table}[h]
  \begin{center}
    \small
    \begin{tabularx}{\linewidth}{Xrrrrrrrr}
      \toprule
    
      & \multicolumn{8}{c}{\textbf{Parameter}} \\
       \cmidrule{2-9}
       
      & $\beta$
      & $T_{\rm inc}$
      & $T_{\rm inf}$
      & $T_{\rm recov}$
      & $T_{\rm fatal}$
      & $P_{\rm fatal}$
      & $E_{0}$
      & $I_{0}$ \\

      \midrule

      True Value
      & 0.25
      & 5.10
      & 6.60
      & 14.00
      & 10.00
      & 0.03
      & 1.00
      & 1.00 \\

      Search Space
      & \newinterval{0}{1}
      & \newinterval{1}{100}
      & \newinterval{1}{100}
      & \newinterval{1}{100}
      & \newinterval{1}{100}
      & \newinterval{0}{1}
      & \newinterval{0}{5}
      & \newinterval{0}{5} \\

      MCMC Proposal Variance
      & 0.10
      & 4.00
      & 4.00
      & 4.00
      & 4.00
      & 0.01
      & 0.50
      & 0.50 \\
      \bottomrule
    \end{tabularx}
    \caption{Parameters used for experiments in \autoref{sec:nonid}}
    \label{tab:param-ranges}
  \end{center}
\end{table}

\begin{table}[h]
  \begin{center}
    \small
    \begin{tabular}{lrrrrr}
      \toprule
      & \multicolumn{5}{c}{\textbf{Hyper Parameter}} \\
      \cmidrule{2-6}
      & ${\rm R}_{\rm 0}$
      & ${\rm A}_{\rm 0}$
      & ${\rm D}_{\rm 0}$
      & N
      & Horizon \\

    \midrule      

      Value
      & 0
      & 5
      & 0
      & 10$^{7}$
      & 400 days \\
      \bottomrule
    \end{tabular}
    \caption{Hyperparameters used to generate the synthetic dataset.}
    \label{tab:param-ranges-hyper}
  \end{center}
\end{table}

\subsection{Discussion on Practical Identifiability Intervals}
\label{sec:A4}

We briefly outline the  connection between the Bayesian and the loss function-based notions of practical identifiability intervals in ~\autoref{sec:notions} for clarity. 

\begin{enumerate}

\item For a strictly convex Lipschitz continuous real-valued function $\mathscr{L}_D:\bTheta \mapsto \R$, with a global minimum $c^*$, its level sets $\{\btheta \in \bTheta |  \mathscr{L}_D(\btheta)  \leq c\}$  for $c \in [c^*, \infty)$ are (possibly  multi-dimensional) convex regions nested within each other~\citep{boyd2004convex}.

\item Given a probability distribution $q(\btheta)$ that is non-zero on $\bTheta$, every level set of  
$\mathscr{L}_D(\btheta)$ has a well-defined probability mass resulting in a bijection $Q_{\mathscr{L}} :  [0,1] \mapsto [c^*, \infty)$. In other words,  for any $\alpha \in [0,1]$, there is a unique level and a unique level set of  $\mathscr{L}_D$ with probability mass equal to $\alpha$.

\item For any function $h:\bTheta \mapsto \R$ that can be expressed as 
$h(\btheta) = g(\mathscr{L}_D(\btheta))$ for a strictly monotonically increasing function $g(\cdot)$, the
level set
$\{\btheta \in \bTheta |  \mathscr{L}_D(\btheta)  \leq c\}$ is the same as
$\{\btheta \in \bTheta |  \mathscr{h}(\btheta)  \leq g(c)\}$
for $c \in [c^*, \infty)$.
Due to the mapping between the level sets, the uniqueness of regions with probability mass equaling $\alpha$ holds true even for $h(\cdot)$.

\item When $\mathscr{L}_D(\btheta)$ is the sum of the negative log-likelihood and appropriate regularization terms corresponding to the prior on $\btheta$, then the posterior distribution $p_D(\btheta)$ is strictly monotonically decreasing with respect to  $\mathscr{L}_D(\btheta)$.

\item With the above definition of
$\mathscr{L}_D(\btheta)$, since 
$-p_D(\btheta)$ turns out to be a strictly monotonically increasing function of $\mathscr{L}_D(\btheta)$, 
the level sets map to each other.
Hence, for a given $\alpha \in [0,1]$, it is the same unique convex region that corresponds to a level set of $\mathscr{L}_D$ and the $\alpha$-HPDI (highest posterior density interval) of $p_D(\btheta)$.

\item Since  $\mathscr{L}_D(\cdot)$ is strictly convex, the level sets corresponding to  parameter-wise loss or profile likelihood $\mathscr{L}^i_D(\cdot)$ are contiguous intervals that are the just the 1-D projections of the level sets of  $\mathscr{L}_D(\cdot)$  along the $i^{th}$ dimension for the same level value .

\item On the other hand, the projection of the credible intervals from $p_D(\btheta)$ on to the $i^{th}$ dimension do not exactly correspond to that of the marginal posterior $p_D^i(\theta_i)$. For $\alpha$ close to 1, these intervals are, however, approximately equal as reflected in the empirical analysis with $\alpha =0.95$.

\end{enumerate}
\end{document}